\documentclass[conference]{IEEEtran}
\IEEEoverridecommandlockouts
\usepackage{cite}
\usepackage{amsmath,amssymb,amsfonts}
\usepackage{algorithmic}
\usepackage{graphicx}
\usepackage{textcomp}
\usepackage{xcolor}
\usepackage{booktabs}
\usepackage[switch]{lineno}
\usepackage{listings}
\usepackage[utf8]{inputenc}
\usepackage{xcolor}
\usepackage{tcolorbox}
\usepackage{hyperref}
\usepackage{subcaption}
\usepackage{csquotes}
\usepackage{multirow}
\usepackage[normalem]{ulem}
\usepackage{xcolor}
\usepackage{todonotes}
\usepackage{subcaption}
\usepackage{caption}
\usepackage{lipsum}
\usepackage{float}
\usepackage{subfloat}
\usepackage{pgfplots, pgfplotstable}
\usepackage{tikz}
\pgfplotsset{compat=1.16}
\usepackage{pgfplotstable}
\usepackage{pgf-pie}
\usepackage{tcolorbox}
\usepackage{ragged2e}
\usepackage{adjustbox} 
\usepackage{ifthen}
\usepackage{soul}
\usepackage{lipsum}
\usepackage{helvet} 
\usepackage{courier} 
\usepackage{inconsolata} 

\usepackage[numbers]{natbib}
\usepackage{balance}

\definecolor{codegreen}{rgb}{0,0.4,0}
\definecolor{codegray}{rgb}{0.5,0.5,0.5}
\definecolor{codepurple}{rgb}{0.58,0,0.82}
\definecolor{backcolour}{rgb}{1,1,0.97}
\definecolor{whitecolour}{rgb}{1,1,1}

\lstdefinestyle{mystyle}{
    backgroundcolor=\color{backcolour},   
    commentstyle=\color{codegreen},
    keywordstyle=\color{magenta},
    numberstyle=\tiny\color{codegray},
    stringstyle=\color{codepurple},
    basicstyle=\ttfamily\footnotesize,
    breakatwhitespace=false,         
    breaklines=true,                 
    captionpos=b,                    
    keepspaces=true,                 
    numbers=left,                    
    numbersep=5pt,                  
    showspaces=false,                
    showstringspaces=false,
    showtabs=false,                  
    tabsize=2
}

\newif\ifpienumberinlegend
\pgfkeys{/number in legend/.code=
    \expandafter\let\expandafter\ifpienumberinlegend
    \csname if#1\endcsname
    \ifpienumberinlegend

    \def\beforenumber##1\afternumber{}%
    \fi,
    /number in legend/.default=true
}

\newboolean{showcomments}
\setboolean{showcomments}{true} 

\ifthenelse{\boolean{showcomments}}
{
  \newcommand{\nbc}[3]{
    \colorbox{#3}{\bfseries\sffamily\scriptsize\textcolor{white}{#1}}
    {\textcolor{#3}{\sf\small$\blacktriangleright$\textit{#2}$\blacktriangleleft$}}
  }
}
{
  \newcommand{\nbc}[3]{}
}

\useunder{\uline}{\ul}{}
\def\BibTeX{{\rm B\kern-.05em{\sc i\kern-.025em b}\kern-.08em
    T\kern-.1667em\lower.7ex\hbox{E}\kern-.125emX}}

\begin{document}

\title{\texttt{AUTOGENICS:} Automated Generation of Context-Aware Inline Comments for Code Snippets on Programming Q\&A Sites Using LLM}

\author{\IEEEauthorblockN {Suborno Deb Bappon, Saikat Mondal, Banani Roy}
\IEEEauthorblockA{Department of Computer Science, University of Saskatchewan, Canada\\ }
{\{suborno.deb, saikat.mondal, banani.roy\}}@usask.ca
}



\maketitle

\begin{abstract}

Inline comments in the source code facilitate easy comprehension, reusability, and enhanced readability. However, code snippets in answers on Q\&A sites like Stack Overflow (SO) often lack comments because answerers volunteer their time and often skip comments or explanations due to time constraints. Existing studies show that these online code examples are difficult to read and understand, making it difficult for developers (especially novices) to use them correctly and leading to misuse. Given these challenges, we introduced \texttt{AUTOGENICS}, a tool designed to integrate with SO to generate effective inline comments for code snippets in SO answers exploiting large language models (LLMs).
Our contributions are threefold.
First, we randomly select 400 answer code snippets (200 Python + 200 Java) from SO and generate inline comments for them using LLMs (e.g., Gemini). We then manually evaluate these comments' effectiveness using four key metrics: accuracy, adequacy, conciseness, and usefulness. Overall, LLMs demonstrate promising effectiveness in generating inline comments for SO answer code snippets.
Second, we surveyed 14 active SO users to perceive the effectiveness of these inline comments. The survey results are consistent with our previous manual evaluation.
However, according to our evaluation, LLMs-generated comments are less effective for shorter code snippets and sometimes produce noisy comments. Third, to address the gaps, we introduced \texttt{AUTOGENICS} that extracts additional context from question texts and generates context-aware inline comments. It also optimizes comments by removing noise (e.g., comments in import statements and variable declarations).
We evaluate the effectiveness of \texttt{AUTOGENICS}-generated comments using the same four metrics that outperform those of standard LLMs. 
\texttt{AUTOGENICS} might (a) enhance code comprehension with context-aware inline comments, (b) save time, and improve developers' ability to learn and reuse code more accurately.

\end{abstract}


\section{Introduction}
\label{sec:introduction}


\begin{figure}[t]
	\centering
	\includegraphics[width=3in]{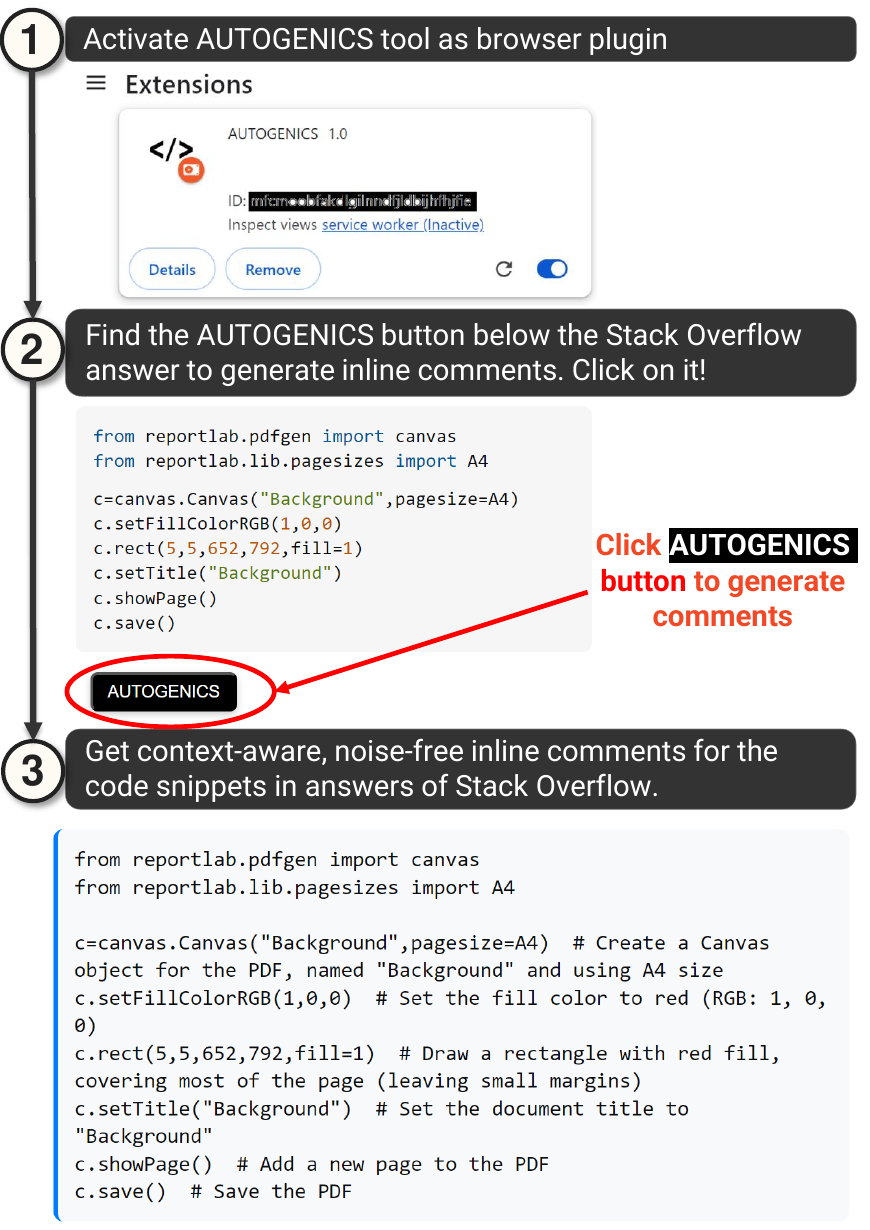}
	\caption{An overview of the \texttt{AUTOGENICS} workflow.}
	\label{fig:editex-workflow}
 \vspace{-3mm}
\end{figure}

Source code comments are essential for enhancing code comprehension, facilitating software maintenance, and promoting reusability \cite{wong2013autocomment, wong2015clocom, zhang2020retrieval, panichella2015would, vinz2008improving, storey2007programmers, kadar2016code, huang2006runtime}.
One key issue with code snippets found in answers on programming Q\&A sites, such as Stack Overflow (SO), is the lack of comments.
%
%
Two factors that might contribute to skip commenting on code snippets are the voluntary nature of SO participation and developers' time constraints \cite{huang2014model, huang2020does, huang2019learning}.
%
%
Additionally, the manual process of adding comments to code is tedious, discouraging developers from doing so \cite{chen2018neural, liang2018automatic}.
Previous research shows that SO code examples often need better usability (e.g., readability) \cite{mondal2023subjectivity}.
Answerers often do not provide adequate explanations for their code \cite{mondal2023subjectivity}. However, undocumented code is a major source of developer frustration, as developers often get confused while reading code snippets without proper comments \cite{ford2015exploring, hu2022practitioners}.
More than $582K$ answer code snippets have neither inline comments nor explanations \cite{api}. 
Such evidence raises severe concerns about correctly reusing the code examples, potentially leading to misuse. Therefore, there is a growing demand for automated tools to generate comments for code snippets \cite{iyer2016summarizing, hu2018deep, alon2018code2seq, hu2020deep, shido2019automatic, leclair2019neural, moore2019convolutional}.
%

A few studies focus on method-level documentation to summarize the purpose and functionalities of methods \cite{wang2020reinforcement, pascarella2019classifying, gao2019neural, hu2018summarizing}. However, method-level comment generation techniques might not be suitable for SO code snippets, as these snippets often consist of a few bare statements rather than complete methods \cite{mondal2022reproducibility}.
%
%
Therefore, inline comments are more suitable in our target context. They clarify the functionalities of each line of code, making it easier to read and understand. Given these points, an investigation is warranted to find a better way to generate effective inline comments for SO answer code snippets that are frequently shared and reused. As far as we know, such investigation has yet to be addressed in existing literature.

Recent advancements in AI, especially LLMs, have revolutionized NLP tasks and achieved state-of-the-art results. LLMs with proper in-context learning and adequate prompts demonstrate superior performance in diverse software development tasks, including documentation \cite{geng2024large, zhao2024automatic, shin2023prompt, zhang2022coditt5}.
In this study, we thus leverage the power of LLMs to generate inline comments on the code snippets included with SO answers. 
We randomly select 400 code snippets (200 Java + 200 Python) found in SO answers. First, we generate inline comments for the code snippets using \texttt{Gemini 1.5 Pro}, which is free. 
We focus solely on standalone code snippets, excluding context (e.g., question descriptions), to evaluate how effectively the language model can generate comments.
Then, we manually evaluate four key metrics - accuracy, adequacy, conciseness, and usefulness - to ensure the effectiveness of the generated inline comments. 
We randomly selected 20 samples and then generated inline comments using \texttt{GPT-4} \cite{gpt4} to see how consistent the comments were across different LLMs.
We further conducted a user study to hear from the practitioners about the effectiveness of the LLM-generated comments. Fourteen professional software developers (who are also active users of SO) participated in this survey.
As a practical outcome of our research, we introduced \texttt{AUTOGENICS}, a tool specifically designed to integrate with SO. 
Figure \ref{fig:editex-workflow} presents the workflow of \texttt{AUTOGENICS}, which involves three straightforward steps: (1) activating the \texttt{AUTOGENICS} tool as a browser plugin, (2) locating the \enquote{AUTOGENICS} button below the code snippets in SO answers, and (3) generating inline code comments by clicking \enquote{AUTOGENICS}.
\texttt{AUTOGENICS} can generate inline comments by considering the context of the questions. It also optimizes comments by removing noise, making it a valuable assistance for software developers.

In this study, we answer three research questions and thus make three major contributions.

\noindent\textbf{RQ1. How effective are LLMs at generating inline comments for code snippets found in Stack Overflow answers?}
This research question aims to evaluate the capability of LLMs (e.g., \texttt{Gemini}) to improve code comprehension by generating effective inline comments for SO answer code snippets. We evaluate the effectiveness of the LLM-generated inline comments by measuring their accuracy, adequacy, conciseness, and usefulness. We employ a 5-point Likert scale to quantify these metrics.
Overall, LLMs demonstrate promising effectiveness. The accuracy, adequacy, and usefulness of LLM-generated inline comments improve with longer code snippets. The performance of the \texttt{Gemini 1.5 Pro} model closely matches that of \texttt{GPT-4}.
%


\noindent\textbf{RQ2. How do developers perceive the effectiveness of these inline comments? Are they interested in an automated tool to generate them?}
Understanding how developers perceive the effectiveness of inline comments is crucial for enhancing the usability and impact of automated tools. We randomly selected eight example code snippets (four Java + four Python) with inline comments and asked the participants to evaluate their effectiveness using the same four metrics as in RQ1.
Our user study confirms the results from RQ1. Approximately 79\% of participants showed strong interest in an automated tool for generating inline comments.

\noindent\textbf{RQ3. Can we introduce a tool into Stack Overflow to generate context-aware, noise-free, inline comments for code snippets? Are these comments more effective than those from standard LLMs?}
We explore the feasibility of introducing tool support to enhance code readability and understanding by automating the generation of context-aware inline comments for code snippets in SO answers. We then introduce \texttt{AUTOGENICS} to assist developers. \texttt{AUTOGENICS} can be easily integrated with the SO site to annotate code snippets with context-aware, noise-free inline comments, outperforming those generated by standard LLMs.

\smallskip
\noindent\textbf {Replication Package} available in our online appendix \cite{replicationPackage}.


\begin{figure}
    \centering
    
    \subfloat[An answer where users requested an explanation of the code to determine its correctness.]
    {\includegraphics[width=3.4in]{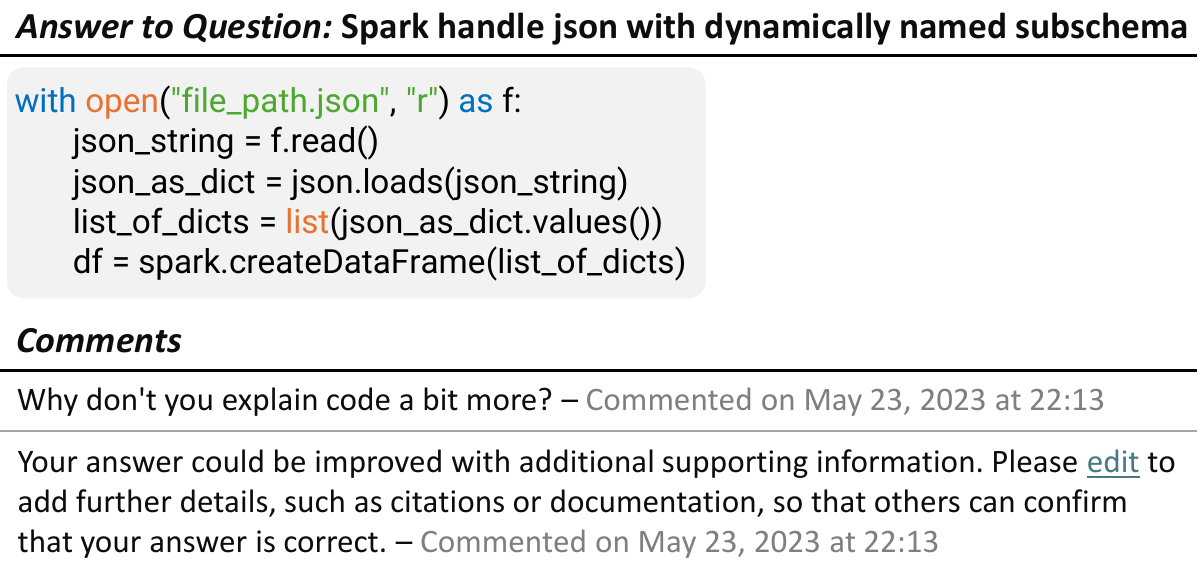}\label{subfig:intro-code}}

    \subfloat[Answer after adding comments using \texttt{AUTOGENICS}.]
    {\includegraphics[width=3.4in]{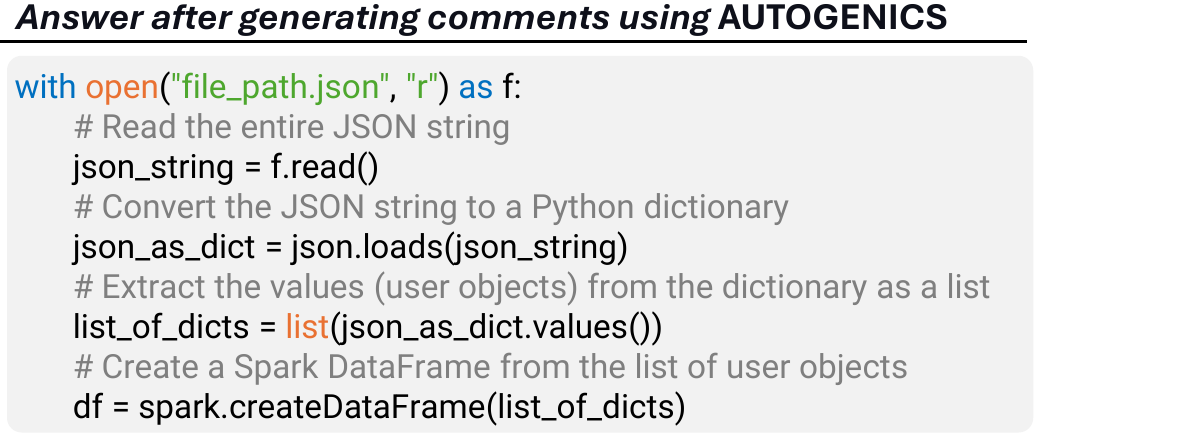}\label{subfig:code-after-comment}}


    \caption{A motivational example~\cite{spark} where users requested a code explanation to validate its accuracy contrasted with the same answer improved by \texttt{AUTOGENICS}-generated comments.}
    \label{fig:motivational-example}
    \vspace{-3mm}
\end{figure}

\begin{figure*}[h]
    \centering
    \includegraphics[width=6in]{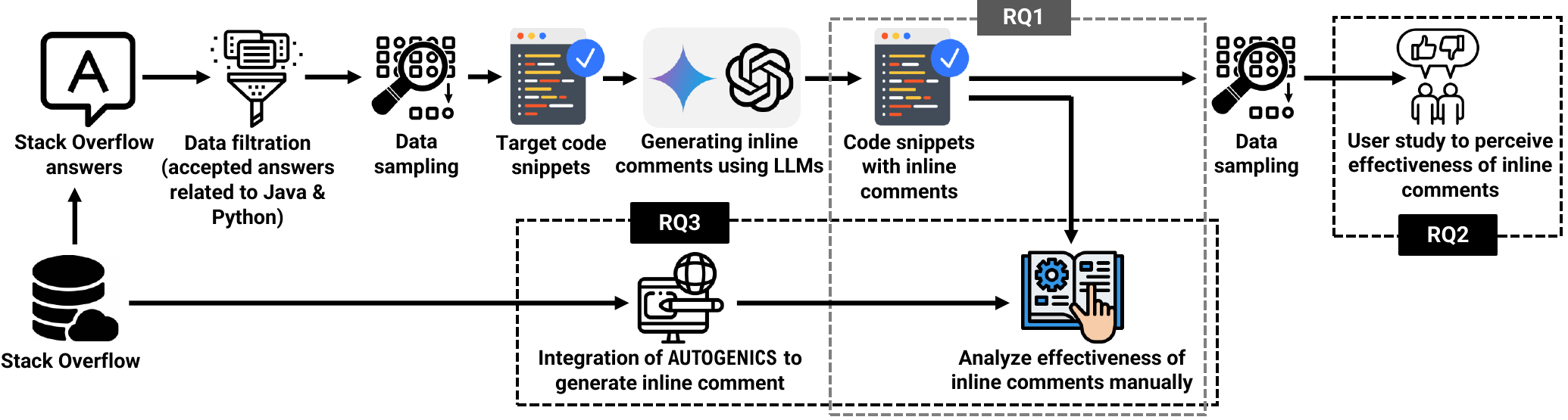}
    \caption{Research methodology for human-centric evaluation of inline code comment generation.}
    \label{fig:methodology}
\end{figure*}

\begin{table*}[!htb]
\centering
\caption{Summary of our dataset \small{(\textbf{Q1-Q4}: \textit{Quartiles 1-4)}.}}
\label{table:total-data-summary}
\resizebox{6.8in}{!}{%
\begin{tabular}{@{}lccccccc@{}}
\toprule
       & \multicolumn{1}{c}{\textbf{\begin{tabular}[c]{@{}c@{}}Total \\ Answers\end{tabular}}} & \multicolumn{1}{c}{\textbf{\begin{tabular}[c]{@{}c@{}}Accepted Answers\\ with Single \\ Code Fragment\end{tabular}}} & \multicolumn{1}{c}{\textbf{\begin{tabular}[c]{@{}c@{}}Q1\\ (1\textless{}= LOC\textless{}= 2)\end{tabular}}} & \multicolumn{1}{c}{\textbf{\begin{tabular}[c]{@{}c@{}}Q2\\ (3\textless{}= LOC\textless{}= 7)\end{tabular}}} & \multicolumn{1}{c}{\textbf{\begin{tabular}[c]{@{}c@{}}Q3\\ (Python: 8\textless{}= LOC\textless{}= 14,\\ Java: 8\textless{}= LOC\textless{}= 16)\end{tabular}}} & \multicolumn{1}{c}{\textbf{\begin{tabular}[c]{@{}c@{}}Q4\\ (Python: 15\textless{}= LOC\textless{}= 695,\\ Java: 17\textless{}= LOC\textless{}= 997)\end{tabular}}} & \multicolumn{1}{c}{\textbf{\begin{tabular}[c]{@{}c@{}}Sampled\\ Answers\end{tabular}}} \\ \midrule
\textbf{Python} & 1606,298                                                                               & 470,393                                                                                                               & 121,065                                                                                                      & 135,144                                                                                                                   & 101,011                                                                                                                                                                     & 112,302                                                                                                                                                                          & 200                                                                                    \\
\textbf{Java}   & 1357,200                                                                               & 393,684                                                                                                               & 98,883                                                                                                       & 108,164                                                                                                                   & 93,513                                                                                                                                                                      & 92,258                                                                                                                                                                           & 200                                                                                    \\ \midrule
\textbf{Total}  & 2963,498                                                                               & 864,077                                                                                                               & 219,948                                                                                                      & 243,308                                                                                                                   & 194,524                                                                                                                                                                     & 204,560                                                                                                                                                                          & 400                                                                                    \\ \bottomrule
\end{tabular}
}
\vspace{-3mm}
\end{table*}

\section{Motivational Example}
\label{sec:motivational-example}

Code segments submitted as part of answers on SO often lack inline comments and are not always adequately explained \cite{mondal2023subjectivity}. Let us consider the example answer in Fig. \ref{subfig:intro-code}. It contains only a Python code snippet. However, the answerer provided neither an explanation nor any comments. As a result, users who were trying to use it to resolve their problems struggled to understand it. A lack of understandability can lead to the misuse of code snippets and the introduction of latent bugs in production code. One user thus commented, ``Why don't you explain code a bit more?''. Another user requested documentation to determine if the answer was correct. Unfortunately, the answer score is zero, which means it is non-useful.

On the other hand, consider the same code snippet with inline comments generated by our tool \texttt{AUTOGENICS}, as shown in Fig. \ref{subfig:code-after-comment}. Each line of code is clearly explained, revealing its function and purpose. These comments make the code easy to understand, even for novice developers. Such comments significantly (a) reduce the time and effort needed to reuse the solution correctly and (b) enhance developers programming knowledge.

This example is one of many that motivates our study. In this study, we attempt to generate inline code comments by introducing tool support. Our tool can (a) assist millions of SO users in generating comments for code snippets in SO answers and (b) enhance the readability and understandability of code.

\section{Study Methodology}
\label{sec:study-methodology}


Fig. \ref{fig:methodology} shows the schematic diagram of our study.
We first collect about three million answers from SO. We collect these answers from two popular programming languages employing one restriction. We then randomly select 400 answers (200 Java + 200 Python) and generate inline comments utilizing LLMs. Then, we manually evaluate their quality from different aspects using four popular metrics.
Second, we study 14 practitioners, as their opinions are crucial in determining the effectiveness of the inline comments.   
Finally, we introduce a tool that can be integrated with the SO site to generate context-aware inline comments and assess their quality manually. 
The following sections discuss different steps of our methodology.

\subsection{Dataset Preparation}

\noindent\textbf{Data Collection.}
Table \ref{table:total-data-summary} lists the dataset for our study.  
We collected a total of 2963,498 answers from SO using StackExchange Data API \cite{api}, all of which were posted on or before February 2024.
In particular, we collected these answers to questions related to two widely used programming languages, Python and Java, to generate inline comments using LLMs. This choice also allows us to assess LLMs' capability to generate comments for various programming language types, including static and dynamic ones. We choose answers under a restriction: the answer must be accepted by the question's owner and contain one code snippet. Accepted answers are widely regarded as reliable solutions. Focusing on one code snippet ensures consistency and avoids potential complications from merging multiple examples. After that, we get 470,393 Python answers and 393,684 Java answers. Finally, we extract code snippets using specialized HTML tags such as \texttt{<code>} under \texttt{<pre>}.

\noindent\textbf{Quartile Analysis.}
We calculate the number of lines (i.e., LOC) of the code snippets and divide them into four quartiles. A quartile divides a dataset into four equal parts, each with 25\% of the data samples. This approach allows us to evaluate the effectiveness of LLMs across code snippets with varying LOC. Table \ref{table:total-data-summary} shows the quartile-wise distribution of our code snippets. Interestingly, the number of code snippets across four quartiles for  Python and Java are evenly distributed.

\noindent\textbf{Data Sampling.}
Our dataset is meticulously prepared, ensuring fairness and representativeness. We randomly select 50 code snippets from each quartile, resulting in a final dataset of 400 code snippets, 200 from each Python and Java. Note that this sample size is statistically significant with a 95\% confidence level and a 5\% error margin \cite{boslaugh2012statistics, mondal2024can}.

\subsection{Inline Comments Generation Using LLMs}
\label{subsec:methodology-inline-comment-generation}

We primarily utilize Google's \texttt{Gemini 1.5. Pro} \cite{gemini} to generate inline comments for our selected code snippets. 
We opted for the \texttt{Gemini} model for two key reasons - (1) it is freely accessible (offers 50 requests per day), thereby enabling easy replication of our findings, and (2)it demonstrates high performance in several software development-related tasks (e.g., code generation) compared to other state-of-the-art models like \texttt{GPT-4} \cite{gptvsgemini}.
According to our analysis, most code snippets in accepted answers lack classes or methods. However, a few of them include these structural elements. LLM models often generate both inline comments and method/class-level documentation for these code fragments. Surprisingly, sometimes, they modify the original snippets. Additionally, some code snippets may contain inline comments. 
Careful prompt design is crucial for retaining the original code snippets and adding inline comments (when necessary). Therefore, we consider these factors when designing effective prompts to ensure the precise generation of inline comments.
Consider the following example prompt that was passed to \texttt{Gemini} to generate inline comments.
The prompt includes three clear instructions with the target code snippet - (1) the purpose of inline code comments, (2) the direction not to alter the given code snippets, and (3) guidance to avoid class/method-level documentation (if there is any).

\begin{table}[H]
    \centering
    \resizebox{3.4in}{!}{%
    \begin{tabular}{p{8cm}}
    \hline\\
    \vspace{-3mm}
    {\fontsize{8}{8.5}\selectfont \ttfamily \noindent \textbf{Inline comment generation prompt (Gemini):} Given the following code snippet: \{CODE\_SNIPPET\}. Generate inline comments to explain what each part of the code does. An inline comment is a single-line comment typically used to explain or clarify a specific line of code. It starts with // for Java and \# for Python. Ensure that you only add inline comments and do not alter the existing code. Avoid adding any comments at the class or method level; focus only on inline comments.}\\
    \hline
    \end{tabular}
    }
    
    \label{table:prompt-1}
\end{table}


In addition to \texttt{Gemini}, we conducted a case study with \texttt{GPT-4} to generate inline comments for a subset of 40 randomly selected code snippets (20 Python + 20 Java evenly distributed across quartiles). We used the same prompt as Gemini to ensure a fair evaluation. This approach allows us to explore \texttt{GPT-4}'s capability to produce inline comments.
Please note that we do not consider question contexts in this phase while generating inline comments.

\subsection{Effectiveness Evaluation of Inline Comments}

\begin{table}[!htbp]
\centering
\caption{Inline comment evaluation metrics.}
\label{tab:metrics-def}
\resizebox{3.4in}{!}{%
\begin{tabular}{lp{7.2cm}}
\toprule
\textbf{Metric} & \textbf{Description}                                                                                                      \\ \midrule
\multirow{2}{*}{\textbf{Accuracy}}                            & The extent to which the generated inline comments correctly describe the functionality and behavior of the code.          \\
\multirow{2}{*}{\textbf{Adequacy}}                            & The degree to which the inline comments provide sufficient information to understand the code without unnecessary details. \\
\multirow{2}{*}{\textbf{Conciseness}}                         & The measure of how brief and to-the-point the inline comments are, avoiding unnecessary verbosity.                        \\
\multirow{2}{*}{\textbf{Usefulness}}                          & The overall helpfulness of the inline comments in enhancing the comprehension and maintenance of the code.                 \\ \bottomrule
\end{tabular}
}
\vspace{-2mm}
\end{table}

Two authors, one with more than seven years and another with over 14 years of professional software development experience in Python and Java, manually evaluate the effectiveness of the inline comments generated by the LLMs.
Previous studies have emphasized the importance of human evaluation for code comments, arguing that automatic metrics like BLEU and ROUGE may not always be reliable \cite{Hu2022Correlating, evtikhiev2023out, wong2013autocomment, Dam2023Enriching, Kovalchuk2022Human, Hashtroudi2023Automated}. Additionally, the lack of ground truth for inline comments on SO code snippets restricts us from utilizing such automated metrics.


We evaluate effectiveness using four metrics: accuracy, adequacy, conciseness, and usefulness (see descriptions Table \ref{tab:metrics-def}) \cite{wong2013autocomment}. We use a 5-point Likert-scale \cite{jebb2021review, nemoto2014likert} to quantify the assessment of each of the metrics. A higher rating signifies better quality of inline comments and vice versa.
Initially, we conducted multiple interactive sessions to discuss and reach a consensus on the evaluation metrics. We then randomly select 80 code snippets with inline comments (40 Python + 40 Java, ten from each quartile) from our selected 400 code snippets. Next, we meticulously evaluate inline comments and assign rankings (1-5) based on the four metrics.
We categorize ranks 1-3 as low and 4-5 as high. We then measure the agreement using Cohen's Kappa \cite{cohen1968weighted, cohen1960coefficient}. The value of {\large $\kappa$} was $0.94$, which means the strength of the agreement is almost perfect. Next, we resolve the remaining few disagreements by discussion. However, the agreement level indicates that any coder can do the rest of the ranking without introducing individual bias. Thus, the first author of this paper evaluates and ranks the remaining samples. We spent a total of 100 person-hours for this manual evaluation.

\begin{table*}[!htbp]
\centering
\caption{Experience, profession, and SO usage frequency of participants.}
\label{tab:survey-participants}
\resizebox{5.5in}{!}{%
\begin{tabular}{@{}cllll|cll|clll@{}}
\hline
\multicolumn{5}{c|}{\textbf{Development Experience (Years)}}                                                                                             & \multicolumn{3}{c|}{\textbf{Professsion}}                                                           & \multicolumn{4}{c}{\textbf{Frequency of SO Usage}}                                                                  \\ \hline
\multicolumn{1}{l}{\textbf{\textless{}= 2}} & \textbf{3-5}               & \textbf{6-10}              & \textbf{11-15}        & \textbf{\textgreater 15} & \multicolumn{1}{l}{\textbf{SW Developer}} & \textbf{Academician}      & \textbf{Student}            & \multicolumn{1}{l}{\textbf{Daily}} & \textbf{Weekly}            & \textbf{Monthly}           & \textbf{Rarely}       \\ \hline
-                                           & \multicolumn{1}{c}{57.1\%} & \multicolumn{1}{c}{42.9\%} & \multicolumn{1}{c}{-} & \multicolumn{1}{c|}{-}   & 64.3\%                                    & \multicolumn{1}{c}{7.1\%} & \multicolumn{1}{c|}{28.6\%} & 64.3\%                             & \multicolumn{1}{c}{21.4\%} & \multicolumn{1}{c}{14.3\%} & \multicolumn{1}{c}{-} \\ \hline
\end{tabular}
}
\end{table*}

\subsection{Developers' perception of the effectiveness of inline comments} 


\begin{figure*}[!htbp]
    \centering
    \includegraphics[width=5.3in]{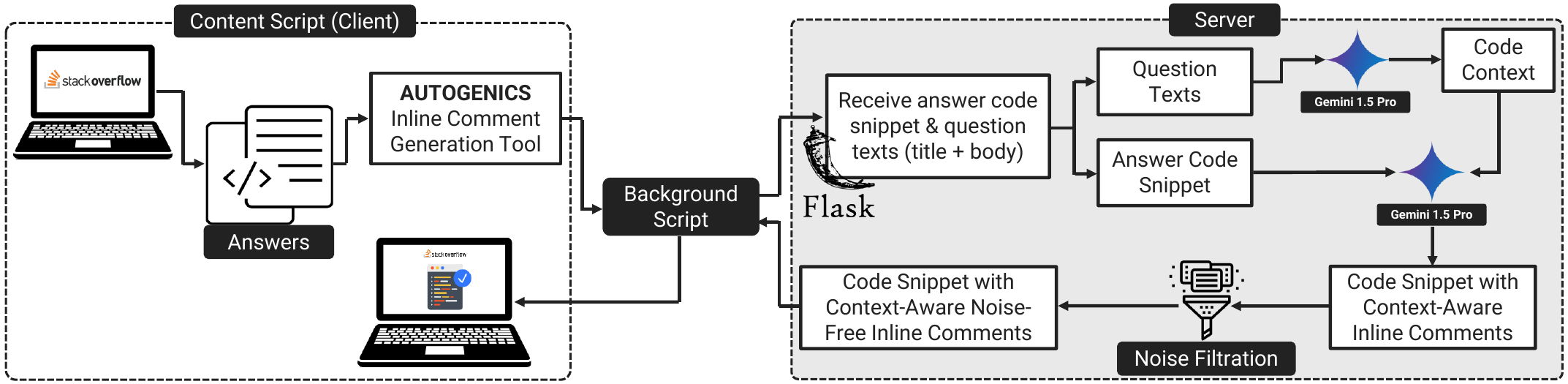}
    \caption{An overview of \texttt{AUTOGENICS} system architecture.}
    \label{fig:autogenics_arch}
    \vspace{-3mm}
\end{figure*}

\noindent\textbf{Survey Design.}
We survey software practitioners to hear how they perceive the effectiveness of the LLM-generated inline comments.
We follow the guidelines and steps outlined by Kitchenham and Pfleeger~\cite{kitchenham2008personal} for conducting personal opinion surveys. We also consider ethical issues from the established best practices~\cite{groves2009survey, singer2002ethical}. For instance, we obtained participants' consent, ensured the confidentiality of their information, and explained the purpose of the survey beforehand. We conducted a pilot survey with three practitioners to gather feedback on (a) the survey's duration and (b) the clarity and comprehensibility of the assigned tasks. Based on their feedback, we made minor modifications and finalized the survey. We inform participants that the estimated time to complete the survey is 15-20 minutes. The responses from the pilot survey were excluded from the final analysis.
Our survey comprises the following five parts.

\noindent\textbf{Consent and Prerequisite.} To be eligible for the survey, participants must confirm their consent, agree to data processing, be familiar with the SO, and have experience in programming languages (e.g., Python or Java).

\noindent\textbf{Participants Information.} In this section, we collect participants' demographic and professional software development experience.

\noindent\textbf{Evaluation of Inline Comments.} Participants are presented with four randomly selected code snippets (Python or Java) with inline comments, one from each quartile generated by standard \texttt{Gemini}. They were asked to evaluate the accuracy, adequacy, conciseness, and usefulness of the inline comments. We provide five options -  very bad (1), bad (2), average (3), good (4) and excellent (5).

\noindent\textbf{Tool Support Needs and Preferences.} In this section, we inquired about participants' interest in automated tools for generating inline code comments for code snippets in SO answers with five options: not interested at all, not very interested, neutral, somewhat interested, and very interested. We also asked their preference on which types of tools they preferred: web app, browser plugin, IDE extension, or API service.


\noindent\textbf{Participants.}
We recruit active users of SO as participants (Table \ref{tab:survey-participants}) and select them as follows.

    $\bullet$ \emph{Snowball Approach:}
    We use convenience sampling to bootstrap the snowball \cite{stratton2021population}. In a collaborative effort, we first contacted a few software developers who were known to us, easily reachable, and working in software companies worldwide. We explained our study goals and shared the online survey with them. We then adopted a snowballing method \cite{bi2021accessibility} to disseminate the survey to several of their colleagues with similar experiences.
     
    $\bullet$ \emph{Open Circular:}
     We circulate the survey to specialized Facebook groups. In particular, we target the groups where professional Python/Java developers discuss their programming problems. We also use LinkedIn to find potential participants because it is one of the largest professional networks.

We recruited 14 participants from countries worldwide (e.g., Canada, Bangladesh) with diverse professions and experience levels and received 14 valid responses (nine for Python + five for Java). Table \ref{tab:survey-participants} summarizes the participants' experience and professions. We then analyze the responses with appropriate tools and techniques based on the question types.

\subsection{\texttt{AUTOGENICS}: Automated Generation of Inline Code Comments} 
\label{subsec:tool-integration}

During the manual evaluation of inline comments (in RQ1), we identified two issues: (1) LLMs frequently generate comments for unnecessary statements in code snippets (e.g., print and import statements), which developers consider as noise \cite{Song2022Do, dau2024docchecker}, and (2) inline comments often fail to represent the purpose of the statements without additional context. We introduce \texttt{AUTOGENICS} that can address these challenges and generate context-aware, noise-free inline comments for SO code snippets.

\noindent\textbf{Architecture of \texttt{AUTOGENICS} and Context-Aware Comment Generation.}
\texttt{AUTOGENICS} is an easy-to-use tool that can be integrated with the user's browser, enabling direct interaction with the SO Q\&A interface. Figure \ref{fig:autogenics_arch} shows the overview of the system architecture of \texttt{AUTOGENICS}. First, we configure the browser extension by defining its properties and permissions. These include permissions to interact with active tabs on SO and to communicate with the local Flask server \cite{flask}. Then, we specify the background service worker and content script to inject functionality into SO pages.

The content script is injected into SO pages to interact with the DOM elements. It locates code snippets within answers and enables a button labeled \enquote{AUTOGENICS} next to each code snippet. The script extracts question texts (title + body) to support LLMs with additional context. 
Upon clicking the `AUTOGENICS' button, the content script sends the candidate code snippet and question texts to the background script. The background script facilitates communication between the content script and the Flask server. 
It listens for messages from the content script and, upon receiving a request, sends the code snippet and question texts to the Flask server. Subsequently, it waits for the server's response, which includes the generated comments.


This tool's main functionality is achieved by running a local Flask server, which acts as the backend for handling requests from the background script. We configure the server to support Cross-Origin Resource Sharing (CORS) to enable interaction with the SO domain. This setup is crucial because \texttt{AUTOGENICS} needs to send requests from a web page to the local server.
Then, we integrate the LangChain framework \cite{langchain} with Google's \texttt{Gemini 1.5 Pro} model. This setup involves loading environment variables, including the Google API key, which is necessary for authenticating requests to the model. 
First, the tool extracts additional code context from SO question texts (title and body). To achieve this, we employ a specially designed prompt with question texts within LangChain as follows.




\begin{table}[H]
    \centering
    \resizebox{3.4in}{!}{%
    \begin{tabular}{p{8cm}}
    \hline\\
    \vspace{-3mm}
    {\fontsize{9}{10}\selectfont \ttfamily \noindent \textbf{Question context extraction prompt (AUTOGENICS):} Extract the main context and key points from the following question description of SO: \{QUESTION\_DESCRIPTION\}. This will help understand the purpose and requirements of the code provided in the answers.}\\
    \hline
    \end{tabular}
    }
    
    \label{table:prompt-1}
\end{table}


Then \texttt{Gemini} processes question texts as input and extracts critical information from them. Later, when combined with code snippets, this contextual information helps better understand the purpose and requirements of the code snippet. This context supports \texttt{AUTOGENICS} to generate context-aware inline comments. To generate context-aware inline comments, we designed the prompt as follows.



\begin{table}[H]
    \centering
    \resizebox{3.4in}{!}{%
    \begin{tabular}{p{9cm}}
    \hline\\
    \vspace{-3mm}
    {\fontsize{9}{10}\selectfont \ttfamily \noindent \textbf{Context-Aware Inline comment generation prompt (AUTOGENICS):}Generate inline comments for the following code snippet: \{CODE\_SNIPPET\}, considering the provided question context:  \{CODE\_CONTEXT\}. An inline comment is a single-line comment typically used to explain or clarify a specific line of code. It starts with // for Java and \# for Python. Ensure that you only generate inline comments and do not alter the existing code. Avoid adding any comments at the class or method level; focus only on inline comments.}\\
    \hline
    \end{tabular}
    }
    
    \label{table:prompt-1}
\end{table}


Once the code snippets with comments are received, the background script sends them back to the content script. Upon receipt, the content script inserts them directly into the SO page immediately after the corresponding code snippet within the answer. The comments are presented in a visually distinctive manner, ensuring they are easily readable and recognizable for users aiming to comprehend the code snippet.

\noindent\textbf{Noise Filtering Mechanism.}
To eliminate noisy comments, we implement a filtering mechanism. This mechanism utilizes regular expressions to detect and remove inline comments from statements that are often treated as noise \cite{noise1, noise2}. Such statements include basic import statements, function definitions, control flow statements, and other typical code patterns that typically do not require explanations (please refer to Table \ref{tab:noisy_patterns} for the list of considered patterns). If any specified patterns are matched, we check whether that statement has an inline comment. If found, we discard that inline comment.

\noindent\textbf{Effectiveness Evaluation of \texttt{AUTOGENICS}-generated comments.}
We randomly select 40 code snippets (20 Python and 20 Java, five from each quartile) from our selected 400 snippets. We generate inline comments for these snippets using \texttt{AUTOGENICS} and manually evaluate their effectiveness using the same four metrics as in RQ1.

\begin{table}[!htbp]
\centering
\caption{Frequently observed statement patterns to filter out noisy inline comments for Java and Java.}
\label{tab:noisy_patterns}
\resizebox{3.4in}{!}{%
\begin{tabular}{p{3.8cm}p{5.8cm}}
\toprule
\textbf{Statement Group} & \textbf{Patterns}                                         \\ \midrule
\multirow{2}{*}{Basic import and print statements}            & print(), import, from {[}module{]} import, System.out.print, System.out.println \\ \hline
Function and class definitions               & def, class, public class, private class, protected class                        \\ \hline
Access modifiers                             & public, private, protected                                                      \\ \hline
Common control flow keywords                 & return, break, continue                                                         \\ \hline
Control structures                           & if, for, while, else, elif, switch, case, default                               \\ \hline
Variable declarations                        & var, let, const, int, float, double, String, boolean                            \\ \bottomrule
\end{tabular}
}
\vspace{-3mm}
\end{table}



\section{Effectiveness Evaluation of LLMs-Generated Inline Comments (RQ1)}
\label{seubsec:rq1-findings}

\begin{table}[!htbp]
\caption{Evaluation summary of Gemini-generated inline comments (\small {\textbf{M} = \textit{Mean}, \textbf{Med} = \textit{Median}; \textbf{Q1-Q4:} \textit{Quartiles 1-4})}.}
\label{table:evaluation-summary-gemini-rq1}
\centering
\resizebox{3.4in}{!}{%
\begin{tabular}{@{}cccccccccc@{}}
\toprule
\multirow{2}{*}{\textbf{Language}} & \multicolumn{1}{c}{\multirow{2}{*}{\textbf{Quartile}}} & \multicolumn{2}{c}{\textbf{Accuracy}}                                   & \multicolumn{2}{c}{\textbf{Adequacy}}                                   & \multicolumn{2}{c}{\textbf{Conciseness}}                                    & \multicolumn{2}{c}{\textbf{Usefulness}}                                     \\ \cmidrule(l){3-10} 
                                   & \multicolumn{1}{c}{}                                   & \multicolumn{1}{c}{\textbf{M}} & \multicolumn{1}{c}{\textbf{Med}} & \multicolumn{1}{c}{\textbf{M}} & \multicolumn{1}{c}{\textbf{Med}} & \multicolumn{1}{c}{\textbf{M}} & \multicolumn{1}{c}{\textbf{Med}} & \multicolumn{1}{c}{\textbf{M}} & \multicolumn{1}{c}{\textbf{Med}} \\ \midrule
\multirow{4}{*}{\textbf{Python}}              & \textbf{Q1}                                           & 4.65                               & 5                                   & 4.12                               & 4                                   & 4.36                               & 4                                   & 3.94                               & 4                                   \\
                                   & \textbf{Q2}                                           & 4.70                               & 5                                   & 4.16                                 & 4                                   & 4.26                               & 4                                   & 4.40                               & 4                                   \\
                                   & \textbf{Q3}                                           & 4.88                               & 5                                   & 4.24                               & 4                                   & 4.14                               & 4                                   & 4.60                               & 5                                   \\
                                   & \textbf{Q4}                                           & 4.74                               & 5                                   & 4.34                               & 4                                   & 3.92                               & 4                                   & 4.76                               & 5                                   \\ \midrule

\multirow{4}{*}{\textbf{Java}}            & \textbf{Q1}                       & 4.80                               & 5                                   & 4.14                               & 4                                   & 4.40                               & 4                                   & 3.86                               & 4                                   \\
                                   & \textbf{Q2}                       & 4.82                               & 5                                   & 4.26                               & 4                                   & 4.32                               & 4                                   & 4.36                                 & 4                                   \\
                                   & \textbf{Q3}                       & 4.86                               & 5                                   & 4.32                               & 4                                   & 4.24                               & 4                                   & 4.56                               & 5                                   \\
                                   & \textbf{Q4}                       & 4.70                               & 5                                   & 4.38                               & 4                                   & 3.96                               & 4                                   & 4.72                               & 5                                   \\ \bottomrule
                                   
\end{tabular}
}
\end{table}

\begin{table}[!htbp]
\centering
\caption{Evaluation summary of our case study with \texttt{GPT-4} in generating inline comments.} 
\label{table:evaluation-summary-gpt-rq1}
\resizebox{3.4in}{!}{%
\begin{tabular}{cccccccccc}
\toprule
\multirow{2}{*}{\textbf{Language}} & \multicolumn{1}{c}{\multirow{2}{*}{\textbf{Quartile}}} & \multicolumn{2}{c}{\textbf{Accuracy}}                                   & \multicolumn{2}{c}{\textbf{Adequacy}}                                   & \multicolumn{2}{c}{\textbf{Conciseness}}                                & \multicolumn{2}{c}{\textbf{Usefulness}}                                 \\ \cmidrule{3-10} 
                                   & \multicolumn{1}{c}{}                                   & \multicolumn{1}{c}{\textbf{M}} & \multicolumn{1}{c}{\textbf{Med}} & \multicolumn{1}{c}{\textbf{M}} & \multicolumn{1}{c}{\textbf{Med}} & \multicolumn{1}{c}{\textbf{M}} & \multicolumn{1}{c}{\textbf{Med}} & \multicolumn{1}{c}{\textbf{M}} & \multicolumn{1}{c}{\textbf{Med}} \\ \toprule
\multirow{4}{*}{\textbf{Python}}            & \centering \textbf{Q1}                                           & 4.8                               & 5.0                                 & 3.8                               & 4.0                                 & 4.4                               & 4.0                                 & 3.8                               & 4.0                                 \\
                                   & \centering \textbf{Q2}                                           & 4.2                               & 4.0                                 & 3.4                               & 3.0                                 & 4.0                               & 4.0                                 & 3.4                               & 3.0                                 \\
                                   & \centering \textbf{Q3}                                           & 4.4                               & 4.0                                 & 3.6                               & 4.0                                 & 4.0                               & 4.0                                 & 3.6                               & 4.0                                 \\
                                   & \centering \textbf{Q4}                                           & 4.2                               & 4.0                                 & 3.4                               & 3.0                                 & 3.6                               & 4.0                                 & 4.2                               & 4.0                                 \\ \midrule
\multirow{4}{*}{\textbf{Java}}              & \centering \textbf{Q1}                                           & 4.6                               & 5.0                                 & 3.6                               & 4.0                                 & 4.2                               & 4.0                                 & 3.6                               & 4.0                                 \\
                                   & \centering \textbf{Q2}                                           & 4.2                               & 4.0                                 & 3.4                               & 3.0                                 & 4.0                               & 4.0                                 & 3.6                               & 4.0                                 \\
                                   & \centering \textbf{Q3}                                           & 4.4                               & 4.0                                 & 3.4                               & 3.0                                 & 4.0                               & 4.0                                 & 3.8                               & 4.0                                 \\
                                   & \centering \textbf{Q4}                                           & 4.2                               & 4.0                                 & 3.4                               & 3.0                                 & 3.8                               & 4.0                                 & 4.0                               & 4.0                                 \\ \bottomrule
\end{tabular}
}
\end{table}

\begin{table*}[!t]
\centering
\caption{Evaluation summary of inline comments by survey participants \small{(\textbf{1} = \textit{Very Bad}, \textbf{2} = \textit{Bad}, \textbf{3} = \textit{Average}, \textbf{4} = \textit{Good}, \textbf{5} = \textit{Excellent}; \textbf{Q1-Q4}: \textit{Quartiles 1-4})}.}
\label{tab:rq2_survey2}
\resizebox{6.7in}{!}{%
\begin{tabular}{@{}cccccccccccccccccccccc@{}}
\toprule
\multirow{2}{*}{\textbf{Language}} & \multirow{2}{*}{\textbf{Quartile}} & \multicolumn{5}{c}{\textbf{\begin{tabular}[c]{@{}c@{}}Accuracy \\ (\% of Participants)\end{tabular}}} & \multicolumn{5}{c}{\textbf{\begin{tabular}[c]{@{}c@{}}Adequacy \\ (\% of Participants)\end{tabular}}} & \multicolumn{5}{c}{\textbf{\begin{tabular}[c]{@{}c@{}}Conciseness \\ (\% of Participants)\end{tabular}}} & \multicolumn{5}{c}{\textbf{\begin{tabular}[c]{@{}c@{}}Usefulness \\ (\% of Participants)\end{tabular}}} \\ \cmidrule(l){3-22} 
                                   &                                    & \textbf{1}         & \textbf{2}         & \textbf{3}         & \textbf{4}         & \textbf{5}         & \textbf{1}         & \textbf{2}         & \textbf{3}         & \textbf{4}         & \textbf{5}         & \textbf{1}          & \textbf{2}          & \textbf{3}          & \textbf{4}         & \textbf{5}         & \textbf{1}          & \textbf{2}         & \textbf{3}         & \textbf{4}         & \textbf{5}         \\ \midrule
\multirow{4}{*}{\textbf{Python}}            & \textbf{Q1}                                 & -                  & -                  & -                  & 66.7               & 33.3               & -                  & -                  & 66.7               & 33.3               & -                  & -                   & -                   & -                    & 22.2               & 77.8               & -                   & 8.1                & 69.7               & 22.2               & -                  \\
                                   & \textbf{Q2}                                 & -                  & -                  & -                  & 33.3               & 66.7               & -                  & -                  & 33.3               & 66.7               & -                  & -                   & -                   & -                   & 66.7               & 33.3               & -                   & -                  & 33.3               & 66.7               & -                  \\
                                   & \textbf{Q3}                                 & -                  & -                  & -                  & 22.2               & 77.8               & -                  & -                  & -                  & 44.4               & 55.6               & -                   & 33.3                & 66.7                & -                  & -                  & -                   & -                  & -                  & 33.3               & 66.7               \\
                                   & \textbf{Q4}                                 & -                  & -                  & -                  & 55.6               & 44.4               & -                  & -                  & -                  & 22.2               & 77.8               & 22.2                & 66.7                & 11.2                & -                  & -                  & -                   & -                  & -                  & 22.2               & 77.8               \\ \midrule
\multirow{4}{*}{\textbf{Java}}              & \textbf{Q1}                                 & -                  & -                  & -                  & 80                 & 20                 & -                  & -                  & 80                 & 20                 & -                  & -                   & -                   & -                   & 20                 & 80                 & -                   & 10                 & 70                 & 20                 & -                  \\
                                   & \textbf{Q2}                                 & -                  & -                  & -                  & 40                 & 60                 & -                  & -                  & 40                 & 60                 & -                  & -                   & -                   & -                   & 80                 & 20                 & -                   & -                  & 20                 & 80                 & -                  \\
                                   & \textbf{Q3}                                 & -                  & -                  & -                  & 20                 & 80                 & -                  & -                  & -                  & 40                 & 60                 & -                   & 20                  & 60                  & 20                 & -                  & -                   & -                  & -                  & 40                 & 60                 \\
                                   & \textbf{Q4}                                 & -                  & -                  & -                  & 60                 & 40                 & -                  & -                  & -                  & 20                 & 80                 & 40                  & 60                  & -                   & -                  & -                  & -                   & -                  & -                  & 20                 & 80                 \\ \bottomrule
\end{tabular}
}
\vspace{-3mm}
\end{table*}

In this section, we manually analyze the effectiveness of LLMs-generated inline code comments using four popular metrics - accuracy adequacy, conciseness, and usefulness.
Table \ref{table:evaluation-summary-gemini-rq1} summarizes the results when we generate inline comments using the \texttt{Gemini} model. According to our analysis, the accuracy of the inline comments in Python code snippets is consistently high across all quartiles. Mean accuracy ranges from 4.65 (first quartile) to 4.88 (third quartile), with a steady median of five. Interestingly, we see an upward trend in the accuracy when LOC increases from the first to the third quartile. However, the accuracy score goes slightly downward in the fourth quartile. We find similar results for the inline comments in Java code snippets.
Such findings suggest that LLMs (e.g., \texttt{Gemini}) can generate more accurate inline comments for longer code snippets. However, accuracy could decline when the LOC is very high, possibly due to the code's increased complexity and context dependence.

Overall, the adequacy score exceeds four for all quartiles (Python + Java), with a median value of four. We observe a consistent, slight improvement in the adequacy score as the LOC increases. As code length increases, there is more complexity to explain, which might motivate LLMs to provide more detailed inline comments for better understanding. Therefore, we see an opposite trend between conciseness and adequacy. While the median values remain consistent at four, mean scores slightly decrease in the higher quartiles. For example, we observe relatively verbose inline comments within lengthy code snippets. 
However, we observe a similar trend as adequacy when evaluating the usefulness metric.


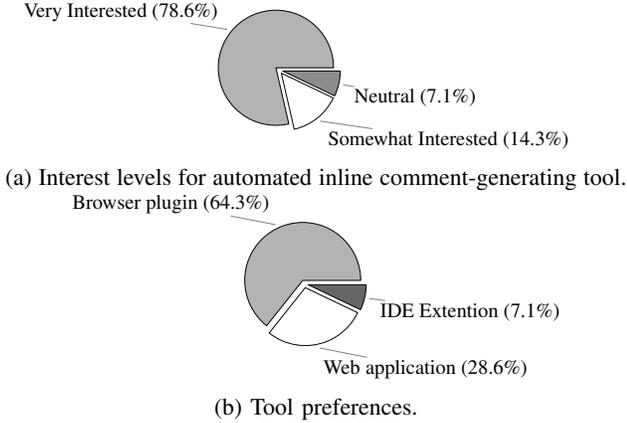
\begin{figure}[!htb]
    \centering
    \subfloat[Interest levels for automated inline comment-generating tool.]
    {
	\resizebox{2.9in}{!}{
        \begin{tikzpicture}
        \pie[explode=0.2, text=pin, number in legend, sum = auto, color={black!30, black!0, black!45, black!60, black!75}]
            { 78.6/{\fontsize{30}{24}\selectfont Very Interested (78.6\%)},
              14.3/{\fontsize{30}{24}\selectfont Somewhat Interested (14.3\%)},
              7.1/{\fontsize{30}{24}\selectfont Neutral (7.1\%)}
            }
        \end{tikzpicture}
        }
        \hspace{4mm}
        \label{subfig:tool-need}
    }
        
    \subfloat[Tool preferences.]
    {
	\resizebox{2.6in}{!}{
        \begin{tikzpicture}
        \pie[explode=0.2, text=pin, number in legend, sum = auto, color={black!30, black!0, black!60, black!75}]
            { 64.3/{\fontsize{30}{24}\selectfont Browser plugin (64.3\%)},
              28.6/{\fontsize{30}{24}\selectfont Web application (28.6\%)},
              7.1/{\fontsize{30}{24}\selectfont IDE Extention (7.1\%)}
            }
        \end{tikzpicture}
        }
        \label{subfig:tool-preference}
        
    }

\caption{Interest levels and preferences for an automated inline comment-generating tool.}
\label{fig:tool-interest-preference}
\vspace{-5mm}
\end{figure}

We conducted a case study to evaluate the performance of the \texttt{GPT-4} model in generating inline comments. We randomly select 40 code snippets from our dataset (20 Python + 20 Java, evenly distributed across quartiles) to generate inline comments utilizing \texttt{GPT-4}. We then manually evaluate the four metrics. Table\ref{table:evaluation-summary-gpt-rq1} summarizes the results of each metric.
The accuracy and conciseness of the \texttt{Gemini 1.5 Pro} model are comparable to those of \texttt{GPT-4}. However, there were notable differences in adequacy and usefulness scores, where \texttt{GPT} underperforms relative to \texttt{Gemini}. Interestingly, \texttt{GPT} generates more accurate inline comments for Python code snippets of shorter lengths (i.e., within the first quartile). Therefore, selecting the \texttt{Gemini 1.5 Pro} model for deploying a tool to generate inline comments is cost-effective and offers performance comparable to or exceeding that of \texttt{GPT-4}.

\begin{tcolorbox}[colframe=black!50, colback=white,left=0pt,right=1pt,top=1pt,bottom=1pt, arc=1pt]
\textbf{Summary RQ1.}  
The \texttt{Gemini 1.5 Pro} model generates highly accurate and adequate inline comments, especially for longer code snippets, while \texttt{GPT-4} excels with shorter Python snippets. Overall, \texttt{Gemini 1.5 Pro} is cost-effective and performs comparably or better than \texttt{GPT-4}, making it a reasonable choice for generating inline comments.
\end{tcolorbox}

\section{Practitioners' Perspective on LLM-Generated Comment Effectiveness \& Tool Preference (RQ2)}

In this section, we survey 14 SO users to perceive the effectiveness of LLM-generated inline comments and tool support preferences for generating them automatically.

\subsection{Effectiveness Evaluation of Inline Comments by Survey Participants.} 
Table \ref{tab:rq2_survey2} shows the effectiveness evaluation summary by the survey participants, which closely aligns with the results of our manual evaluation (Table \ref{table:evaluation-summary-gemini-rq1}). These consistent results enhance the confidence of our evaluation.

The number of participants who rated the comment's accuracy as Excellent increased from the first to the third quartile, followed by a slight decline, similar to what we observed in RQ1. The adequacy of inline comments for the code snippets was primarily rated as Average (3) for the first quartile and Good (4) for the second quartile. However, it was mostly assessed as Excellent for the third and fourth quartiles. For example, 77.8\% of the participants rated the adequacy of inline comments for the fourth quartile's code snippets as Excellent. Similar findings are shown for Usefulness.

On the contrary, the inline comments of the code snippets from the upper two quartiles were evaluated as less concise (e.g., Good or below ratings). The opposite results are found in the lower two quartiles. For example, 77.8\% of the participants rated the inline comments of the code snippet from the first quartile as Excellent (5).

\subsection{Tool Support Needs and Preferences.}
As shown in Fig. \ref{subfig:tool-need}, approximately 79\% of the participants expressed strong interest in tool support for generating inline comments for SO answer code snippets. No participants selected the options `Not Very Interested' or `Not Interested At All'. These results indicate the participants' high demand for an inline comments generation tool.

Fig. \ref{subfig:tool-preference} shows tool supports preference pie chart. For example, 64.3\% of the participants prefer an inline comments generator tool as a browser plugin, while 28.6\% prefer a web application.
Such preference encourages us to consider developing tool support as a browser plugin for generating inline comments.

\begin{table*}[!htbp]
\centering
\caption{Evaluation Summary of conventional and context-aware noise-free inline comments \small{(\textbf{M} = \textit{Mean}, \textbf{Med} = \textit{Median}; \textbf{WO Context} = \textit{Without Context}, \textbf{W Context} = \textit{With Context}; \textbf{Q1-Q4}: \textit{Quartiles 1-4)}.}}
\label{tab:w_wo_con}
\resizebox{6.5in}{!}{%
\begin{tabular}{@{}cccccccccccccccccc@{}}
\toprule
\multirow{3}{*}{\textbf{Language}} & \multirow{3}{*}{\textbf{Quartile}} & \multicolumn{4}{c}{\textbf{Accuracy}}                                            & \multicolumn{4}{c}{\textbf{Adequacy}}                                            & \multicolumn{4}{c}{\textbf{Conciseness}}                                         & \multicolumn{4}{c}{\textbf{Usefulness}}                                          \\ \cmidrule(l){3-18} 
                                   &                                    & \multicolumn{2}{c}{\textbf{WO Context}} & \multicolumn{2}{c}{\textbf{W Context}} & \multicolumn{2}{c}{\textbf{WO Context}} & \multicolumn{2}{c}{\textbf{W Context}} & \multicolumn{2}{c}{\textbf{WO Context}} & \multicolumn{2}{c}{\textbf{W Context}} & \multicolumn{2}{c}{\textbf{WO Context}} & \multicolumn{2}{c}{\textbf{W Context}} \\ \cmidrule(l){3-18} 
                                   &                                    & \textbf{M}        & \textbf{Med}        & \textbf{M}        & \textbf{Med}       & \textbf{M}        & \textbf{Med}        & \textbf{M}        & \textbf{Med}       & \textbf{M}        & \textbf{Med}        & \textbf{M}        & \textbf{Med}       & \textbf{M}        & \textbf{Med}        & \textbf{M}        & \textbf{Med}       \\ \midrule
\multirow{4}{*}{\textbf{Python}}            & \textbf{Q1}                                 & 4.4               & 4.0                 & 4.8               & 5.0                & 3.4               & 3.0                 & 4.2               & 4.0                & 4.4               & 4.0                 & 4.6               & 5.0                & 3.8               & 4.0                 & 4.4               & 4.0                \\
                                   & \textbf{Q2}                                 & 4.4               & 4.0                 & 4.8               & 5.0                & 3.6               & 4.0                 & 4.4               & 4.0                & 4.0               & 4.0                 & 4.4               & 4.0                & 4.2               & 4.0                 & 4.6               & 5.0                \\
                                   & \textbf{Q3}                                 & 4.8               & 5.0                 & 5.0               & 5.0                & 3.8               & 4.0                 & 4.4               & 4.0                & 4.0               & 4.0                 & 4.2               & 4.0                & 4.2               & 4.0                 & 4.6               & 5.0                \\
                                   & \textbf{Q4}                                 & 4.6               & 5.0                 & 4.8               & 5.0                & 4.2               & 4.0                 & 4.6               & 5.0                & 3.8               & 4.0                 & 4.0               & 4.0                & 4.4               & 4.0                 & 4.8               & 5.0                \\ \midrule
\multirow{4}{*}{\textbf{Java}}              & Q1                                 & 4.4               & 4.0                 & 4.8               & 5.0                & 3.8               & 4.0                 & 4.2               & 4.0                & 4.0               & 4.0                 & 4.6               & 5.0                & 3.6               & 4.0                 & 4.2               & 4.0                \\
                                   & \textbf{Q2}                                 & 4.6               & 5.0                 & 5.0               & 5.0                & 3.8               & 4.0                 & 4.4               & 4.0                & 4.0               & 4.0                 & 4.4               & 4.0                & 4.0               & 4.0                 & 4.4               & 4.0                \\
                                   & \textbf{Q3}                                 & 4.6               & 5.0                 & 5.0               & 5.0                & 4.0               & 4.0                 & 4.6               & 5.0                & 3.8               & 4.0                 & 4.4               & 4.0                & 4.4               & 4.0                 & 4.6               & 5.0                \\
                                   & \textbf{Q4}                                 & 4.4               & 4.0                 & 4.8               & 5.0                & 4.2               & 4.0                 & 4.6               & 5.0                & 3.6               & 4.0                 & 4.2               & 4.0                & 4.6               & 5.0                 & 4.8               & 5.0                \\ \bottomrule
\end{tabular}
}
\vspace{-3mm}
\end{table*}

\begin{tcolorbox}[colframe=black!50, colback=white,left=0pt,right=1pt,top=1pt,bottom=1pt, arc=1pt]
\textbf{Summary RQ2.} 
The majority of participants found the LLM-generated inline comments effective and expressed a strong demand for tool support. In particular,  they preferred an automated inline commenting tool as a browser plugin on the SO site.
\end{tcolorbox}

\begin{figure*}[!htb]
    \centering
    \includegraphics[width=6.5in]{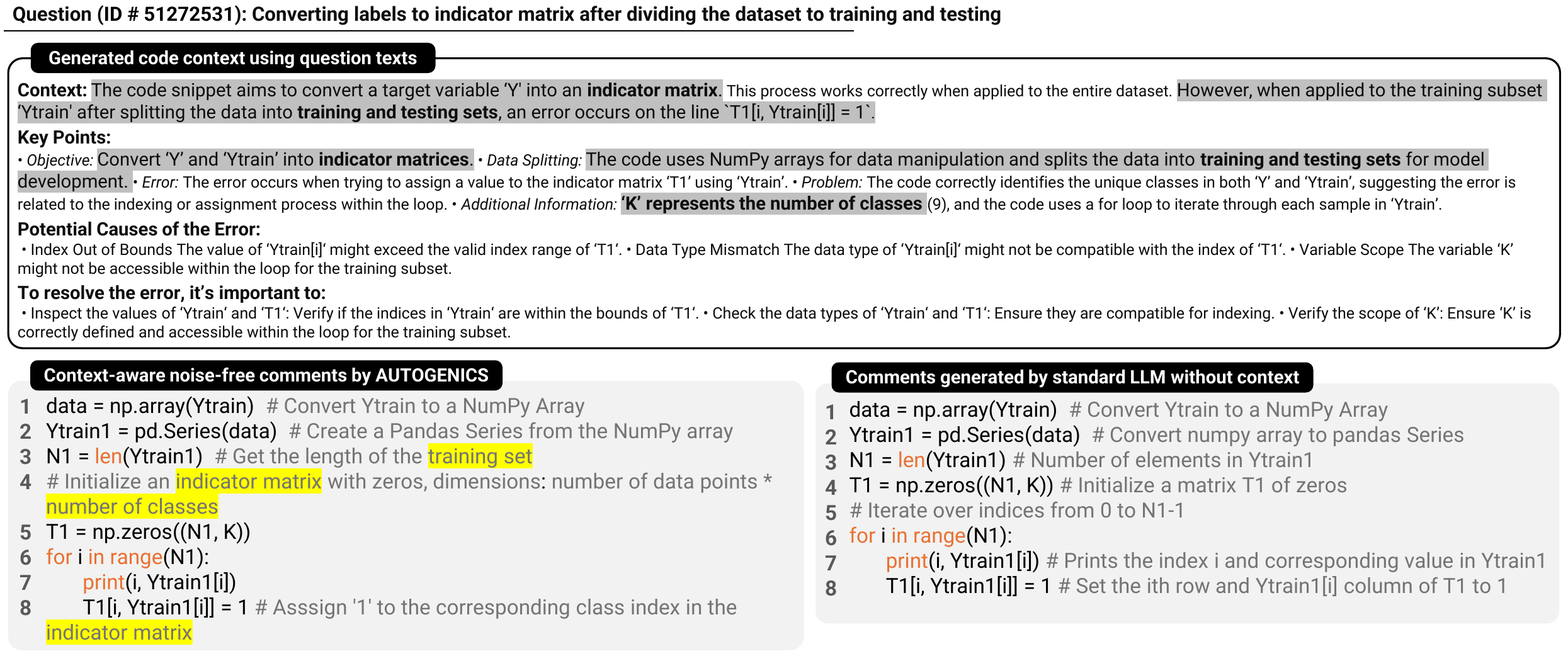}
    \caption{Example of how \texttt{AUTOGENICS} uses context to generate effective inline comments.}
    \label{fig:context-aware-inline-comments}
    \vspace{-3mm}
\end{figure*}

\section{\texttt{AUTOGENICS}: A Browser Plugin to Generate Context-Aware Noise-Free Inline Comments (RQ3)}

In this section, we compare the effectiveness between comments generated without context by \texttt{Gemini} and comments generated with context by the \texttt{AUTOGENICS} tool. \texttt{AUTOGENICS} considers additional code context from question texts and filters out noise. Table \ref{tab:w_wo_con} summarizes the results between standard \texttt{Gemini} and \texttt{AUTOGENICS}.

For Python, the evaluation demonstrates significant enhancements when additional code context is incorporated. For example, scores of ‘Accuracy' without context are consistent across quartiles, ranging from 4.4 to 4.8. In contrast, context-aware scores increase from 4.8 to 5.0, with medians consistently at 5.0.
‘Adequacy' scores without context are moderate, ranging from 3.4 to 4.2, but they improve significantly with context and achieve mean scores ranging from 4.2 to 4.6, with a median of 4.0. These findings highlight the crucial role of contextual information from SO questions in generating relevant comments.
For ‘Conciseness', Python inline comments without context score around 3.8 to 4.4, indicating they are fairly concise. Context-aware comments maintain or slightly improve conciseness, scoring between 4.0 and 4.6 with almost similar median values.
The usefulness of inline comments shows a consistent pattern—scores range from 3.8 to 4.4 without context. However, scores improve significantly (ranging from 4.4 to 4.8) with context, often achieving median scores of 5.0. These results clearly show the advantages of additional code context in generating inline comments to enhance their effectiveness.
Based on our analysis, we observe a similar enhancement in the effectiveness of inline comments for Java code snippets when incorporating contextual information.

Additional context enhances \texttt{AUTOGENICS}'s ability to align inline comments with the intended functionality and logic (see Table \ref{tab:w_wo_con}). Such context improves accuracy by ensuring relevant comments and correctly describing code operations. Understanding context also allows \texttt{AUTOGENICS} to include critical insights that might otherwise be overlooked, ensuring thorough explanations of code complexities. \texttt{AUTOGENICS} also filters out unnecessary distractions. Consider the examples in Fig. \ref{fig:context-aware-inline-comments}, which demonstrate how incorporating context enhances the ability of \texttt{AUTOGENICS} to generate more effective inline comments.
First, \texttt{AUTOGENICS} extract the question texts. It then prompts \texttt{Gemini} with the question texts to produce a structured context.
For example, it produces context, key points, potential causes of the error, and important points to resolve errors for the given question (see Fig. \ref{fig:context-aware-inline-comments}). Next, \texttt{AUTOGENICS} passes the candidate answer code snippets and this context to generate inline comments. 
In this scenario, \texttt{AUTOGENICS} identifies the inherent meaning of \texttt{\textbf{Ytrain1}} as a ‘training set'. It gets this contextual information from the context,  \textbf{\enquote{However, when applied to the training subset `Ytrain' after splitting the data into training and testing sets, an error occurs on the line `T1[i, Ytrain[i]] = 1'.}}. Besides, \texttt{AUTOGENICS} properly pinpoints \texttt{\textbf{T1}} as an indicator matrix and \texttt{\textbf{K}} as the number of classes. It fetches the context for the \texttt{\textbf{T1}} from this line- \textbf{\enquote{The code snippet aims to convert a target variable `Y' into an indicator matrix.}}, and \texttt{\textbf{K}} variable from this line- \textbf{\enquote{`K' represents the number of classes (9), and the code uses a for loop to iterate through each sample in `Ytrain'.}}. In addition, the filtration mechanism of \texttt{AUTOGENICS} filters out comments for the \texttt{\textbf{for}} and \texttt{\textbf{print}} statements, where the standard \texttt{Gemini} model generates comments.

\begin{tcolorbox}[colframe=black!50, colback=white,left=0pt,right=1pt,top=1pt,bottom=1pt, arc=1pt]
\textbf{Summary RQ3.} 
\texttt{AUTOGENICS} is a user-friendly tool designed to be integrated with the SO Q\&A site. It can generate context-aware, noise-free inline comments that significantly improve the overall quality of standard LLM-generated comments.
\end{tcolorbox}

\section{Discussion}
\label{sec:discussion}

In this section, we discuss the key findings and implications of this study.

\subsection{Key Findings}

\noindent\textbf{Consistently High Accuracy Showcases LLM Mastery.}
The LLM's consistently high accuracy ratings for Python and Java demonstrate its robust understanding of programming concepts across different languages and code lengths correctly. LLMs achieve peak comment accuracy in the third quartile of code length, striking the ideal balance between context richness and manageable complexity. Additionally, its performance benefits from incorporating additional code context.



\noindent\textbf{Dilemma of Clarity Vs. Conciseness.}
As code complexity increases, the LLM faces challenges keeping comments concise and often provides more detailed explanations to describe code snippets accurately. Such a scenario reflects the model's effort to balance clarity and thoroughness, emphasizing the importance of capturing the essence of the code without oversimplification through more extensive comments.

\noindent\textbf{Insights into LLM's Context-Driven Commenting Process.}
LLMs prioritize understanding entire code contexts before generating inline comments. Thus, they can generate high-quality comments. 
As shown in Listing~\ref{lst:python_example}, the \texttt{Gemini} generates an inline comment indicating that ‘a' is a PyTorch tensor. 
Such comments involve analyzing common coding patterns and leveraging extensive training data to accurately infer and explain code elements.

\begin{lstlisting}[language=Python, caption=Python code example illustrating the LLM's context-driven inline commenting process., label=lst:python_example]{example.py}
a_n = a.numpy()  # Convert the PyTorch tensor to a NumPy array
# Apply a function along the 2nd axis, summing the powers of 2 of the non-zero elements in each row
a_n = np.apply_along_axis(func1d=lambda x: np.sum(np.power(2,np.where(x==1))[0]), axis=2, arr=a_n) 
a = torch.Tensor(a_n)  # Convert the NumPy array back to a PyTorch tensor

\end{lstlisting}

\subsection{Implications}


Our research findings on automated inline comment generation for SO answer code snippets and the development of \texttt{AUTOGENICS} will provide a basis for future studies. Future \textbf{researchers} can extend our findings to enhance automated code comprehension and documentation capabilities for online code snippets. Additionally, our approach emphasizes the potential of integrating automated tools into coding environments to improve developer productivity and coding efficiency. 


\texttt{AUTOGENICS}, a browser plugin integrated with SO, demonstrates the feasibility and benefits of real-time comment assistance for \textbf{developers} (e.g., SO users) to generate context-aware inline comments in code snippets. It will promote faster comprehension of unfamiliar code, particularly aiding novices in understanding and integrating solutions more accurately.

\texttt{AUTOGENICS} benefits \textbf{educators}, including programming bloggers and tutorial makers, by (a) providing instant feedback on code clarity, (b) ensuring their code examples are well-documented and easy to understand, and thus (c) enhancing educational content. \textbf{Stack Overflow site owners} can integrate \texttt{AUTOGENICS} to improve user experience by enabling real-time inline comments on code snippets, promoting better code readability and documentation.

\section{Threats to Validity}


\emph{External Validity} is related to how broadly our results can be generalized. Our results may not be generalized to all SO answers code snippets. To mitigate this threat, we analyze statistically significant samples from two popular programming languages - Java and Python. Java is statically typed, whereas Python is a dynamically typed programming language. We also categorize code snippets based on LOC and take samples evenly distributed across quartiles. We see that the results from both languages are consistent. Thus, we believe that our insights can be generalized to other programming languages. Moreover, we investigate a wide variety of answers to different types of programming problems in order to combat potential bias in our results. However, we caution readers to refrain from over-generalizing our results. Another threat to generalizability is the use of specific language models. To address this, we conducted a case study with \texttt{GPT-4} in addition to \texttt{Gemini 1.5 Pro}, thereby mitigating the threat.

Threats to \emph{internal validity} relate to experimental errors and biases \cite{tian2014automated}. We manually evaluate the effectiveness of inline comments using four metrics that could introduce bias. However, the agreement between the two annotators was almost perfect (i.e., {\large $\kappa$} $= 0.94$), which ensures the robustness and consistency of our evaluations. We surveyed 14 developers who evaluated the effectiveness of the inline comments. Their results were consistent with ours, further validating our evaluation.

Threats to \emph{construct validity} relate to the suitability of evaluation metrics. To mitigate this threat, we evaluate the effectiveness of the inline comments using four appropriate metrics - accuracy, adequacy, conciseness, and usefulness. \emph{Statistical Conclusion} threats concern the fact that the data is sufficient to support the claims. We considered statistically significant samples in our result analyses.

Snowball sampling relies on referrals and may have a sampling bias. However, we also selected participants using an open circular approach and collected their responses anonymously. Table \ref{tab:survey-participants} shows that our participants have diverse experiences and professions. Such diversity offers validity and applicability to our survey findings.

\section{Related Work}
\label{sec:literature}



Several studies investigate block-level comments, which summarize source code and provide a high-level overview of the purpose and logic of a block of code \cite{sridhara2011automatically, wong2013autocomment, wong2015clocom, iyer2016summarizing, huang2020towards}. Sridhara et al. \cite{sridhara2011automatically} introduced a technique to identify sequences of statements, conditions, and loops in code that could be summarized into higher-level actions. Then, they generated descriptions for these segments using their templates. Wong et al. \cite{wong2013autocomment} developed a method to extract code descriptions from a programming Q\&A site (e.g., SO). Then, they utilized these insights to produce comments for equivalent code segments in open-source projects. Researchers also utilized code clone detection techniques to find and reuse comments from code libraries in open-source software \cite{wong2015clocom}.

The block comment generation domain has also benefited from learning-based approaches, which treat code as text sequences or interpret Abstract Syntax Trees (AST) as sequences. For example, Iyer et al. \cite{iyer2016summarizing} introduced CODE-NN, an LSTM-based neural network model that takes code sequences as input and produces sequences of comment tokens. On the other hand, Huang et al. \cite{huang2020towards} combined heuristic rules with learning-based techniques to develop a reinforcement learning strategy for generating block comments. 

Numerous studies investigate method-level comments, which describe a method's overall intent, parameters, return values, and functionality.
Sridhara et al. \cite{sridhara2010towards} applied the Software Word Usage Model (SWUM) and heuristic-based techniques to select keywords from code. They create templates to explain Java methods. Vassallo et al. \cite{vassallo2014codes} introduced a technique to extract method comments leveraging Q\&A discussion of SO.

Several studies highlight the effectiveness of Abstract Syntax Trees (ASTs) in capturing structural properties in order to improve the quality of method-level comments \cite{hu2018deep, alon2018code2seq, shido2019automatic, hu2020deep, leclair2019neural}.
These techniques largely depend on programming language and language-specific dictionaries.
To address these issues, Moore et al. \cite{moore2019convolutional} designed a CNN model treating code as character sequences to manage dictionary size effectively. Li et al. \cite{li2021secnn} introduced SeCNN, which integrates lexical and syntactic details to improve comment quality and handle longer dependencies in code.

Li et al. \cite{li2022setransformer} introduced SeTransformer, a transformer-based architecture that enhances upon CNNs and RNNs with a self-attention mechanism for simultaneous text and structural feature analysis of code. On the other hand, Yang et al. \cite{yang2021comformer} developed ComFormer, integrating Transformer models with a fusion method to improve comment quality. Meanwhile, Xu et al. \cite{xu2020towards} explored local and global encoders with Graph Attention Networks for contextual information in comment generation. Kuang et al. \cite{kuang2022code} proposed GTrans, combining Graph Neural Networks with Transformers for comprehensive code representation.

The studies mentioned above focus on generating block-level and method-level comments using techniques such as Information Retrieval, templating, or Deep Learning approaches. To our knowledge, our study pioneers the use of LLMs for generating inline comments on code snippets found in SO answers. We demonstrate the effectiveness of these comments through manual analysis and user study for two widely used programming languages. We introduce \texttt{AUTOGENICS}, a tool for generating context-aware, noise-free inline comments. It has the potential to significantly enhance code comprehension and assist millions of developers in effectively utilizing and reusing code resources.

\section{Conclusion}

Inline code comments play a crucial role in enhancing code comprehension, readability, and reusability, particularly in programming Q\&A sites like SO. 
First, we investigate the capability of standard LLMs (e.g., \texttt{Gemini}) to generate effective inline code comments. We randomly select 400 code snippets (200 Python + 200 Java) extracted from accepted answers on SO and employ standard LLMs to generate inline comments. We manually assess four key metrics—accuracy, adequacy, conciseness, and usefulness of the comments. Additionally, we surveyed 14 software developers who are active users of SO to perceive the effectiveness of these inline comments. Our evaluation demonstrates the promise of LLMs in generating inline comments. However, it has a few limitations, such as a lack of effectiveness for shorter code snippets and the presence of noisy comments.
We then introduced \texttt{AUTOGENICS}, a tool leveraging LLM as a browser plugin to automatically generate context-aware, noise-free inline comments for code snippets in SO answers. It can overcome the limitations of standard LLMs by utilizing additional code context from question texts and a noise filtration mechanism. By optimizing comments and removing irrelevant noise, \texttt{AUTOGENICS} aims to improve the overall usability of code snippets on SO, facilitating better learning and reuse practices among developers.

In future studies, we intend to explore the effectiveness of \texttt{AUTOGENICS} across different programming languages and Q\&A platforms. Additionally, we plan to conduct an expert survey to gather user feedback on \texttt{AUTOGENICS}, aiming to enhance its usability based on their insights.

\smallskip
\noindent\textbf{Acknowledgment.}
This research is supported in part by the industry-stream NSERC CREATE in Software Analytics Research (SOAR). 

\bibliographystyle{unsrtnat}

{\small
\bibliography{bibliography}
}

\end{document}